\pdfoutput=1
\documentclass[lettersize,journal]{IEEEtran}
\usepackage{amsmath,amsfonts}
\usepackage{multirow}
\usepackage{algorithmic}
\usepackage{array}
\usepackage[caption=false,font=normalsize,labelfont=sf,textfont=sf]{subfig}
\usepackage{textcomp}
\usepackage{stfloats}
\usepackage{url}
\usepackage{verbatim}
\usepackage{graphicx}
\def\BibTeX{{\rm B\kern-.05em{\sc i\kern-.025em b}\kern-.08em
    T\kern-.1667em\lower.7ex\hbox{E}\kern-.125emX}}
\usepackage{balance}

\usepackage{xcolor}
\definecolor{myBlue}{rgb}{0,0.1,0.9}

\usepackage{xspace}

\usepackage{epstopdf}
\begin{document}

\title{All-rounder: A Flexible AI Accelerator with Diverse Data Format Support and Morphable Structure for Multi-DNN Processing}

\author{Seock-Hwan Noh,~\IEEEmembership{Graduate Student Member,~IEEE;}
        Seungpyo Lee;
        Banseok Shin;
        Sehun Park,~\IEEEmembership{Graduate Student Member,~IEEE;}
        Yongjoo Jang,~\IEEEmembership{Graduate Student Member,~IEEE;}
        and~Jaeha~Kung,~\IEEEmembership{Member,~IEEE}
\IEEEcompsocitemizethanks{\IEEEcompsocthanksitem  S.-H. Noh and S. Park are with the Department of Electrical Engineering and Computer Science, Daegu Gyeongbuk Institute of Science and Technology (DGIST), Daegu, 42988, South Korea. Email: \{nosh3332, aeun3690\}@dgist.ac.kr
\IEEEcompsocthanksitem S. Lee was previously affiliated with DGIST and currently with Fitogether, Seoul, 04323, South Korea. Email: lsp5490@gmail.com
\IEEEcompsocthanksitem B. Shin was previously affiliated with DGIST and currently with Samsung Electronics, Suwon, 16677, South Korea. Email: shin.banseok@gmail.com
\IEEEcompsocthanksitem Y. Jang and J. Kung are with the School of Electrical Engineering, Korea University, Seoul, 02841, South Korea. Email: \{jyjoo, jhkung\}@korea.ac.kr (\textit{J. Kung is the corresponding author})}
\thanks{Manuscript received September 04, 2024; revised December 28, 2024.}
}

\markboth{IEEE Transactions on Very Large Scale Integration (VLSI) Systems}%
{How to Use the IEEEtran \LaTeX \ Templates}

\maketitle

\begin{abstract}
Recognizing the explosive increase in the use of AI-based applications, several industrial companies developed custom ASICs (e.g., Google TPU, IBM RaPiD, Intel NNP-I/NNP-T) and constructed a hyperscale cloud infrastructure with them. These ASICs perform operations of the inference or training process of AI models which are requested by users. 
Since the AI models have different data formats and types of operations, the ASICs need to support diverse data formats and various operation shapes. 
However, the previous ASIC solutions do not or less fulfill these requirements. 
To overcome these limitations, we first present an area-efficient multiplier, named all-in-one multiplier, that supports multiple bit-widths for both integer and floating point data types.
Then, we build a MAC array equipped with these multipliers with multi-format support.
In addition, the MAC array can be partitioned into multiple blocks that can be flexibly fused to support various DNN operation types.
We evaluate the practical effectiveness of the proposed MAC array by making an accelerator out of it, named All-rounder.
According to our evaluation, the proposed all-in-one multiplier occupies 1.49$\times$ smaller area compared to the baselines with dedicated multipliers for each data format.
Then, we compare the performance and energy efficiency of the proposed All-rounder with three different accelerators showing consistent speedup and higher efficiency across various AI benchmarks from vision to LLM-based language tasks.

\end{abstract}

\begin{IEEEkeywords}
Hardware acceleration, deep neural networks, multiple data format support, multi-tenant execution
\end{IEEEkeywords}

\section{Introduction}
\IEEEPARstart{N}{owadays}, AI-based applications are utilized in many fields of our daily lives, such as intelligent self-driving cars, language translation, and content recommendation on over-the-top services. 
However, running inference or training of AI models using single-precision arithmetic, i.e., FP32, which is generally used in CPUs and GPUs, causes expensive computational cost and high runtime due to fetching a large number of parameters and dealing with iterative operations.
To mitigate this burden, various algorithmic techniques for data quantization on activations and/or weights have been explored~\cite{ptq_4bit, parameter8_training, ptq_eccv_8bit, ptq_calibration, bfloat16, inference_8bit_transformer, mixed_training, fp8_training_2022, hybrid_fp8_training, hybrid_fp8_a}. 
Additionally, numerous hardware designs that support quantized data formats have been proposed in both academia and industry for mobile devices~\cite{arm_n78, edge_tpu, edge_bison, edge_blade, exynops_isscc}.

In datacenters, they receive many user requests on various AI applications.
However, each application may operate on an AI model with different bit-widths and architectures that are determined by the application developer~\cite{nestdnn}.
To meet these diverse requirements, companies such as NVIDIA, IBM, Google, and Intel, to name a few, have developed custom ASICs to support various data formats. 
However, these accelerators occupy large area or consume high computing power since they have dedicated multipliers as many as supported bit-precisions in their compute units (e.g., NVIDIA Tensor Cores~\cite{nvidia_h100}, IBM RaPiD~\cite{edge_rapid}), or are unable to support low bit-precision (e.g., Google TPU~\cite{TPU_v4}).
In addition, they have high resource utilization only on specific AI workloads due to lack of hardware flexibility.
For instance, Intel NNP-I~\cite{nnp_i} or NNP-T~\cite{nnp-t} has high resource utilization for compute-centric operations (e.g., channel-wise accumulable convolution), and Google TPU~\cite{TPU_v4} has high utilization for memory-centric operations (e.g., multi-head attention)~\cite{layerweaver}.

Furthermore, a single ASIC chip in the datacenter can receive requests for applications that require simultaneous execution of multiple AI models. 
For instance, a single chip might receive a request for online inference of augmented or virtual reality (AR/VR). 
To deliver a successful service to users, AI models for various tasks, such as object detection, classification, and depth estimation, must be processed concurrently~\cite{xrbench}. 
In addition, a single chip can be shared by multiple AI inference queries to reduce the latency. 
Accordingly, to provide online services for applications that require concurrent execution of various tasks, a single ASIC node needs to handle multiple AI instances simultaneously. 
However, ASICs deployed in datacenters are not optimized for this purpose, as industrial companies have primarily focused on improving performance across computing nodes~\cite{tp_share} or running a single-tenant AI model on an ASIC node as fast as possible~\cite{TPU_v4}.

Considering these limitations, we propose a flexible AI accelerator, named \textit{All-rounder}, which is a versatile hardware solution designed to address these challenges. 
The All-rounder supports multiple data types and precisions with a novel area-efficient multiplier, and it efficiently runs diverse AI operations by maximizing hardware utilization of multiply-and-accumulation (MAC) units through a morphable MAC array that can be fused and split as needed. 
The main contributions of our work are summarized as follows.
\begin{enumerate}
    \item \textbf{Area-efficient all-in-one multiplier:} We developed an area-efficient multiplier capable of supporting exponents ranging from 1 to 8 bits and mantissas of 3 or 7 bits for a floating point (FP) data type. The multiplier can also support signed and unsigned integer (INT) multiplications for 4b or 8b inputs.
    \item \textbf{Morphable MAC array:} We explore limitations of the TPU-like systolic array (SA) and introduce a flexible MAC array to resolve them. The MAC array in All-rounder consists of sub-arrays that can be fused or partitioned based on operation types or multi-tenant scenarios. The proposed MAC array ensures high hardware resource utilization regardless of the number of multi-tenant AI models and types of AI operations.
    \item \textbf{Microarchitecture design and evaluation:} We implemented the full microarchitecture of All-rounder with the proposed multipliers and the morphable array structure. 
    We validate the effectiveness of All-rounder using seven representative AI models with three state-of-the-art AI accelerators and a high-end GPU card.
\end{enumerate}

\section{Motivation}\label{sec:motivation}

\subsection{Area-efficient Multiple Data Format Support}\label{sec:motivation_a_content}
Cloud servers receive requests from users for both training and inference of AI models. 
Since AI algorithms involve a large number of weight parameters and extensive datasets, the inference and training of AI models require a substantial amount of MAC operations and large memory footprint~\cite{zcomp, lightnorm}. 
To alleviate these computational burdens, weights and intermediate features are quantized to low bit-precision. 
In general, data used for inference are quantized to low bit-width INT values, whereas the data used for training are quantized to low bit-width FP numbers~\cite{precision_nature}.

\begin{figure*}[t]
    \centering
    \includegraphics[scale=0.60]{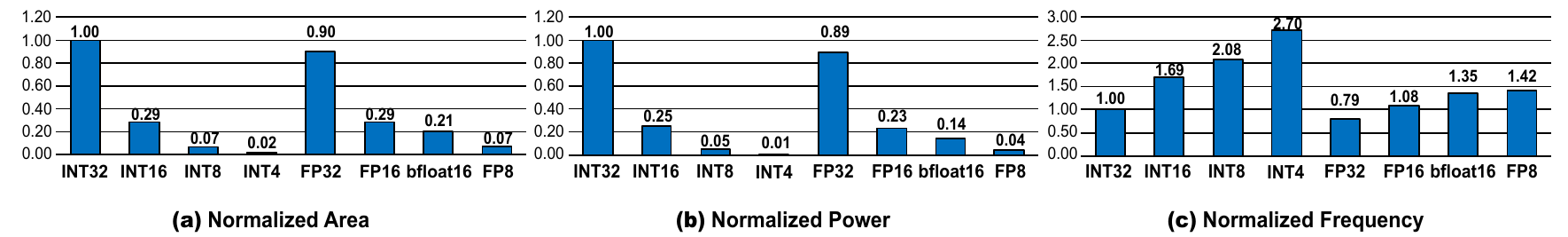}
    \caption{Synthesized results of multipliers for various data types and precisions. They are synthesized in 28nm CMOS technology using Synopsys DesignWare IPs~\cite{dw_synopsys}. (a), (b) and (c) show the area, power consumption, and maximum operating frequency of each multiplier normalized by INT32 counterpart, respectively.}
    \label{fig:Synthesis_result_motivation}        
\end{figure*}

Also, users may request AI services at various data types and bit-precisions depending on their application requirements.
Thus, ASICs designed to support both inference and training must support both INT and FP data types, as well as multiple bit-precisions. 
However, commercial ASICs typically support a limited number of precisions or include dedicated multipliers for each supported data format within their compute units. 
For instance, Google TPU v4 supports bfloat16 and INT8. 
If the TPU is required to handle an AI algorithm quantized to a lower precision than it supports, e.g., FP8 or INT4, it processes operations with higher power consumption than required using bfloat16 or INT8 compute units.
Specifically, Fig.~\ref{fig:Synthesis_result_motivation} shows the synthesized results of FP and INT multipliers across various precisions. When TPU v4 receives a request for an AI algorithm quantized to INT4, the algorithm is processed with an INT8 multiplier, which consumes 3.5$\times$ more power than the INT4 multiplier. 
Additionally, IBM RaPiD supports data formats by incorporating dedicated multipliers for each data format, such as FP16, FP8, INT4, and INT2~\cite{edge_rapid}. 
However, this approach of having dedicated hardware for multiple data types and precisions increases area overhead and leaves some part of MAC units idle since only one precision is normally used in AI operations\footnote{Mixed precision can be employed during the training~\cite{mixed_training, nvidia_mixed_precision}, but a single precision is used in each step of the training. For example, when FP16 and FP32 are exploited to train an AI model, FP16 is utilized for the forward and backward steps, and FP32 is used for the weight update step to accommodate the wide dynamic range of weight gradients.}.

\begin{figure*}[t]
    \centering
    \includegraphics[scale=0.60]{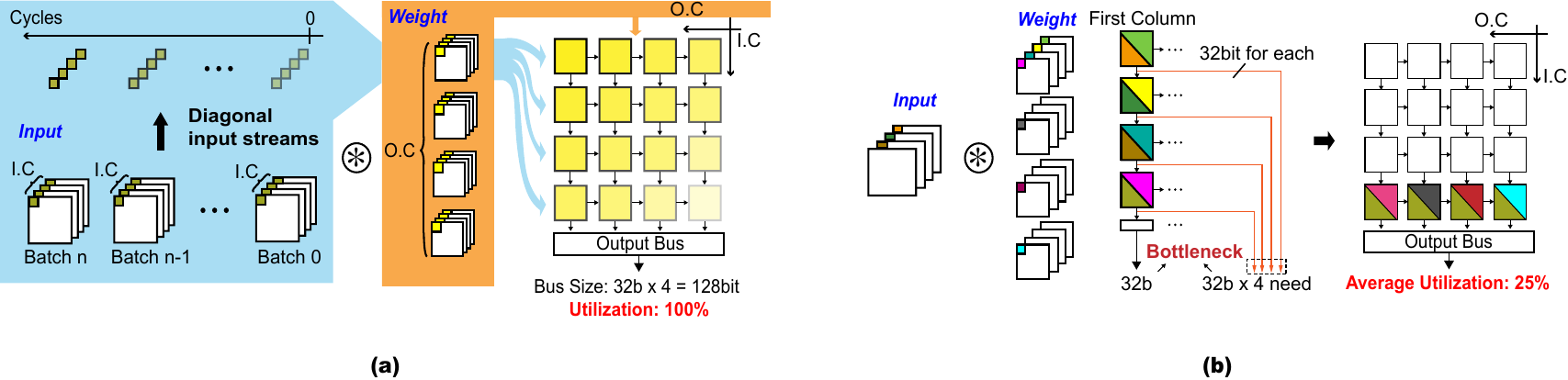}
    \caption{Data mappings on a 4$\times$4 systolic array (SA) for convolution operations commonly used in AI vision tasks. (a) Convolutions where outputs of MAC operations are accumulated across the input channels, and (b) convolutions where results of MAC operations are not accumulated across the input channels. }
    \label{fig:motivation_a}
\end{figure*}

\begin{table}[t]
     \centering
     \caption{Two types of AI operations and their examples}
     \includegraphics[scale=0.70]{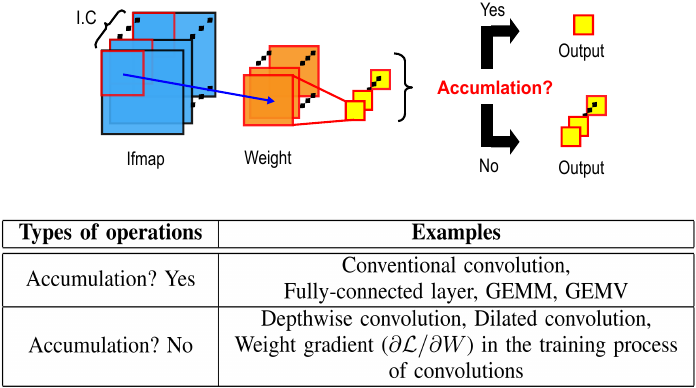}
     \label{tab:classification_of_operations}       
 \end{table}

\begin{figure*}[t]
    \centering
    \includegraphics[scale=0.60]{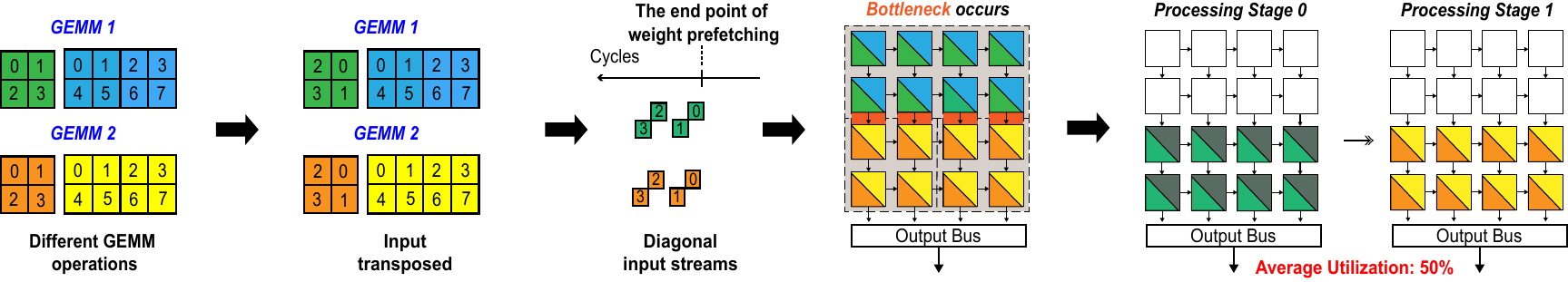}
    \caption{Example of data mapping of multiple AI workloads, i.e., two GEMMs, from two natural language processing models, on the 4$\times$4 systolic array.}
    \label{fig:motivation_b}
\end{figure*}

\subsection{Under-utilization of MAC Units in the Systolic Array}\label{sec:motivation_b_content}
The operations of DNN workloads can be categorized into two types: operations in which results of MAC operations are accumulated in the direction of the input channel and operations in which they are not.
Table~\ref{tab:classification_of_operations} provides examples of these two categories. 
Fig.~\ref{fig:motivation_a} illustrates how data are mapped to a 4$\times$4 TPU-like systolic array (SA) for convolution operations commonly used in AI vision tasks\footnote{If the convolution is remapped into a general multiply matrix (GEMM) using the Im2Col operation, convolution operations can be mapped to SA structures in the form of GEMMs, similar to GPU devices~\cite{sigma}.}. 
When performing convolutions in the SA, data across input channels are mapped to a column of the SA, and thus generating a partial sum for each output channel at each column.
It is noteworthy that the size of output bus, which delivers results to the partial sum buffer, is determined by the number of columns in the SA. 
As depicted in Fig.~\ref{fig:motivation_a}-(a), in the convolution operation that accumulates the results of MAC operations across input channels, partial sums are generated at the bottom of the SA by accumulating all input channel data mapped to SA columns. 
Thus, partial sums streamed out from the SA fully occupy the output bus.
On the contrary, as shown in Fig.~\ref{fig:motivation_a}-(b), mapping data to all MAC units in the operation that does not accumulate results of MAC operations across input channels requires a wider output bus since more partial sums are generated per clock cycle. 
To prevent this, input/weight data must be mapped to the SA so that partial sums do not overfit the output bus. Thus, the convolution that does not accumulate the results of MAC operations in the input channel direction, e.g., depthwise convolution, results in low MAC utilization.

On the other hand, modern natural language processing (NLP) models have attention-based structures~\cite{gpt3, gpt2, llama2}.
The computation of attention involves multiple general matrix multiply (GEMM) operations for key, query, and value generation. 
Thus, ASICs in datacenters need to support GEMM operations with varying matrix sizes and shapes, which are determined by the dimension of embedding vectors, the number of input tokens, and the number of attention heads.
In addition, ASICs need to support multi-tenant execution on a single chip to efficiently respond to requests for multiple AI models.
However, due to the rigid nature of the SA structure, it shows low MAC utilization for multi-tenant execution and GEMM operations with tall/wide matrices. 
Fig.~\ref{fig:motivation_b} illustrates an example showing that two wide GEMM operations cannot be performed simultaneously in the typical SA structure.

\section{Area-efficient All-in-one Multiplier}\label{sec:all_in_one_multiplier}
To enable a single AI accelerator to support both inference and training processes, it is essential to accommodate a variety of data types and precisions. 
To meet this requirement, accelerators deployed in the datacenters of cloud providers already support diverse data formats~\cite{nvidia_h100, TPU_v4, edge_rapid}. 
However, incorporating dedicated multipliers for each supported precision within a compute unit increases area while also leaving parts of the MAC units inactive. 
To address this issue, we have developed an area-efficient multiplier capable of effectively supporting multiple precisions for both FP and INT data types. 
This section details the proposed multiplier implemented in our All-rounder accelerator.

\begin{figure}[t]
    \centering
    \includegraphics[scale=0.60]{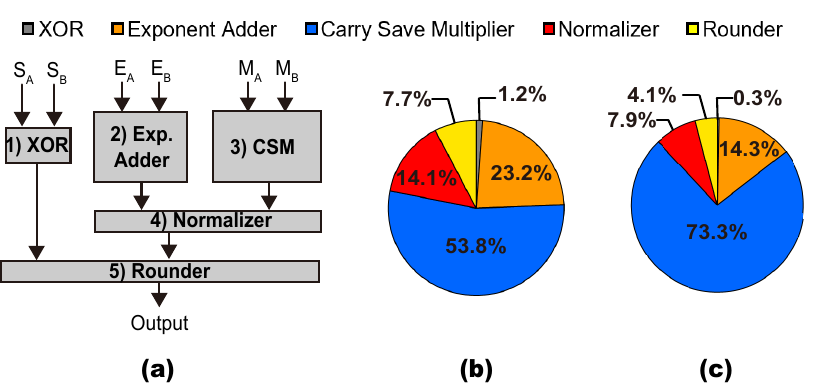}\vspace{-2mm}
    \caption{(a) Structure of a floating point multiplier. (b) Area breakdown of FP8 multiplier. (c) Area breakdown of bfloat16 multiplier.}
\label{fig:fp_multiplier}        
\end{figure}

\textbf{(Background) floating point multiplication:} Multiplication of two floating point numbers $x_A$ = $(-1)^{S_A} \times (1.M_A) \times 2^{E_A + Bias}$ and $x_B$ = $(-1)^{S_B} \times (1.M_B) \times 2^{E_B + Bias}$, where $S$ is the sign, $M$ is the mantissa and $E$ is the exponent of the number $x$, is performed as follows~\cite{fp_multiplier}:

\begin{enumerate}
\item Calculate the sign (i.e., $S_A \oplus S_B$) 
\item Add Exponents (i.e., $E_A + E_B - Bias$)
\item Multiply significands (i.e., $1.M_A \times 1.M_B$)
\item Normalize the result 
\item Round off the normalized result
\end{enumerate}
Fig.~\ref{fig:fp_multiplier}-(a) shows the structure of a basic FP multiplier which each sub-module is numbered with the corresponding multiplication step.

\begin{figure*}[ht]
    \centering
    \includegraphics[scale=0.51]{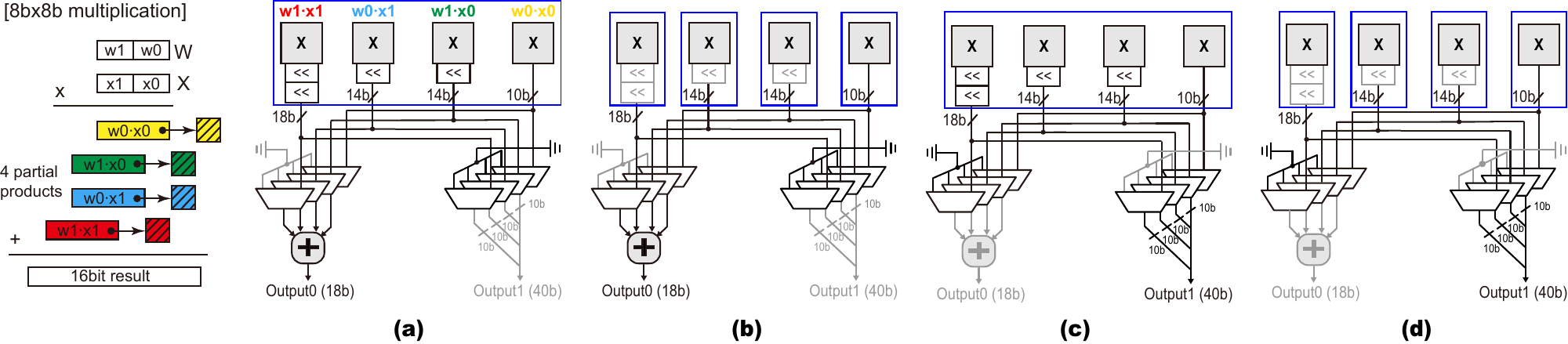}
    \caption{Examples on how the restructured CSM works for some data format combinations: (a) INT8 $\times$ INT8, (b) INT4 $\times$ INT4, (c) bfloat16 $\times$ bfloat16, (d) FP8 $\times$ FP8.}
    \label{fig:carry_save_multiplier}        
\end{figure*}

\textbf{Key ideas: }
Fig.~\ref{fig:fp_multiplier}-(b) and (c) present the area breakdowns of the FP8 \{s:1, e:4, m:3\} and bfloat16 \{s:1, e:8, m:7\} multipliers\footnote{The data format is denoted as \{sign bit, exponent bits, mantissa bits\} throughout this paper.}. 
Notably, carry save multipliers (CSMs) account for the largest area among the sub-modules in each FP multiplier.
We propose an area-efficient, all-in-one multiplier capable of \textbf{i)} supporting various mantissa bit-widths by restructuring the CSM, and \textbf{ii)} multiple bit-widths for the exponent by enabling the exponent adder to manage a programmable bias.
Additionally, \textbf{iii)} the multiplier offers bit-level flexibility for the INT data type by selectively gating all sub-modules except for the CSM. 
As a result, the proposed multiplier efficiently supports both INT and FP data types and bit-widths with minimal area overhead, leveraging a single CSM.

\textbf{Reconstructed carry save multiplier: }
The CSM is an unsigned multiplier responsible for multiplying two significands (i.e., $1.M_A \times 1.M_B$). 
To enable support for multiple mantissa precisions, we replace the CSM with a precision-scalable multiplier. 
For example, if a CSM performs an unsigned multiplication of 8b$\times$8b, it is replaced by four unsigned 4b$\times$4b sub-multipliers.
The reconstructed CSM can perform multiplications of 4b$\times$4b, 4b$\times$8b, 8b$\times$4b, and 8b$\times$8b by selectively shifting and summing the results of the sub-multipliers, similar to~\cite{bitfusion}. 
Moreover, the sub-multipliers within the precision-scalable multiplier are further enhanced to support both signed and unsigned operations. 
Specifically, the unsigned 4b$\times$4b sub-multipliers are replaced with 5b$\times$5b multipliers that include a sign bit.
Fig.~\ref{fig:carry_save_multiplier} shows the structure of the reconstructed CSM and its operation for some precision modes.
The reconstructed CSM can produce one multiplication result in 8b$\times$8b mode, two results in 4b$\times$8b and 8b$\times$4b modes, and four results in 4b$\times$4b mode.
In FP mode, the multiplication results are not summed through the selective adder. 
However, in INT mode, the results are summed, producing a single final output.

\begin{figure}[t]
    \centering
    \includegraphics[scale=0.50]{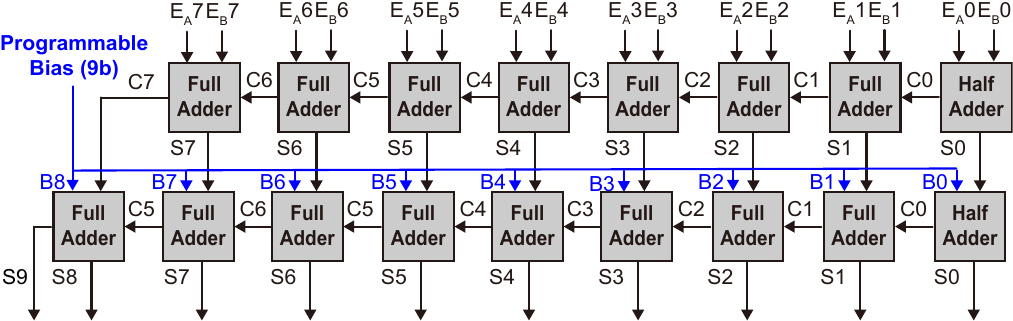}
    \caption{Structure of a programmable exponent adder.}    
    \label{fig:exponent_adder}
\end{figure}

\textbf{Programmable exponent adder: }
The exponent adder in a FP multiplier is composed of two stages of unsigned adders. 
The first stage performs the addition of exponents (i.e., $E_{add} = E_A + E_B$), and the second stage subtracts a bias from the result of the first stage (i.e., $E_{sub} = E_{add} - Bias$). 
conventional FP multipliers support a predefined bias value. 
For example, the exponent adder in a FP32 multiplier supports the bias value of 127~\cite{IEEE754}. 
To support scalable bit-widths for the exponent, it is necessary to allow for different bias values. 
In other words, a programmable bias of $2^{(E.L-1)}-1$, where $E.L$ represents the length of exponent bits, must be supported.
Fig.~\ref{fig:exponent_adder}-(d) illustrates the structure of the proposed exponent adder. 
The multiplier can support exponent bit-widths ranging from 1b to Nb (where N = 8 in the figure), allowing the second stage of the unsigned adder to receive a programmable bias value that aligns with the precision mode of the FP data type.

\begin{figure}[t]
    \centering
    \includegraphics[scale=0.65]
    {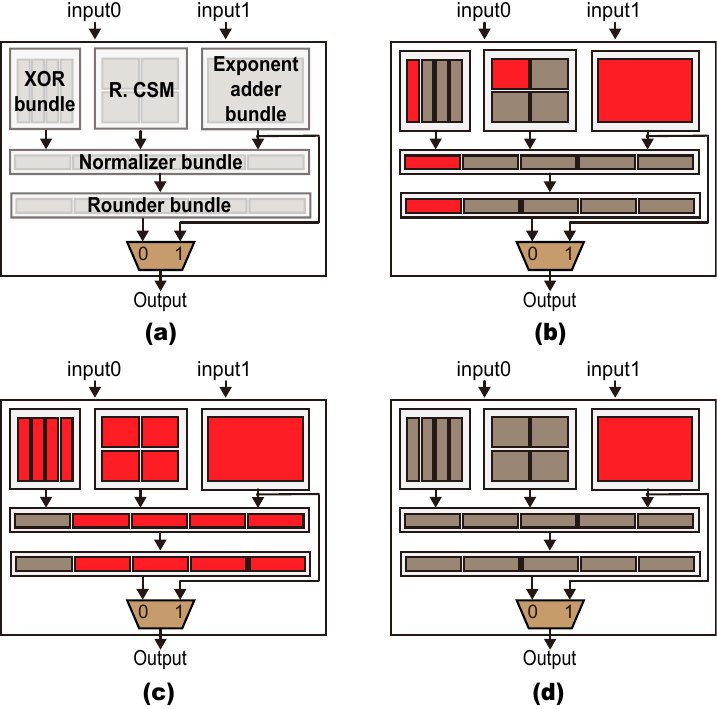}\vspace{-2mm}
    \caption{(a) Overall structure of the proposed multiplier and (b-d) activated sub-modules (colored in red) within the multiplier at different operation modes, i.e., bfloat16, FP8 and INT8, respectively.}
    \label{fig:multiplier_examples}
\end{figure}

\textbf{Overall structure of proposed multiplier:}Fig.~\ref{fig:multiplier_examples}-(a) presents the overall structure of our proposed multiplier. 
If the reconstructed CSM supports multiplication of 4b and 8b combinations, as shown in Fig.~\ref{fig:carry_save_multiplier}, the reconstructed CSM produces one output in 8b$\times$8b mode and four outputs in 4b$\times$4b mode. 
Since the maximum number of outputs from the CSM is four, the multiplier is designed with an XOR bundle and an exponent adder bundle, each comprising four XOR logics and four programmable exponent adders.
The results from the reconstructed CSM are then normalized by combining the outputs from the exponent adder bundle. 
To handle the combined outputs, which have varying bit-widths depending on the precision mode, a normalizer bundle is used. 
This bundle includes five normalizers, i.e., one for the single output in 8b$\times$8b mode and four for the outputs in 4b$\times$4b mode, ensuring proper normalization across all supported FP modes.
Followed by the normalization, the results are rounded to the selected bit precision.
To align with the varying number of outputs from the normalier bundle, the multiplier includes rounding logics as many as the number of normalizers.
These rounding logics are collectively referred to as the rounder bundle.


\textbf{Examples of operating modes:} The proposed multiplier can support exponents in FP operations with exponent bit-widths ranging from 1b to 8b, as well as mantissa precisions of 3b and 7b (i.e., 4b and 8b significands). 
Furthermore, for INT data types, the multiplier supports both signed and unsigned multiplication in 4b and 8b configurations. 
Fig.~\ref{fig:multiplier_examples}-(b-d) depict how sub-modules of the proposed multiplier are activated in the bfloat16, FP8, and INT8 mode, respectively.
In bfloat16 mode, except for the CSM, only one sub-module in each bundle becomes active. 
In FP8 mode, four sub-modules in the bundle work in parallel to perform four FP8 multiplications. 
In the case of INT8 mode, all sub-modules except for the CSM are gated, allowing the output of the CSM to directly serve as an output of the multiplier.

\begin{figure*}[t]
    \centering
    \includegraphics[scale=0.49]{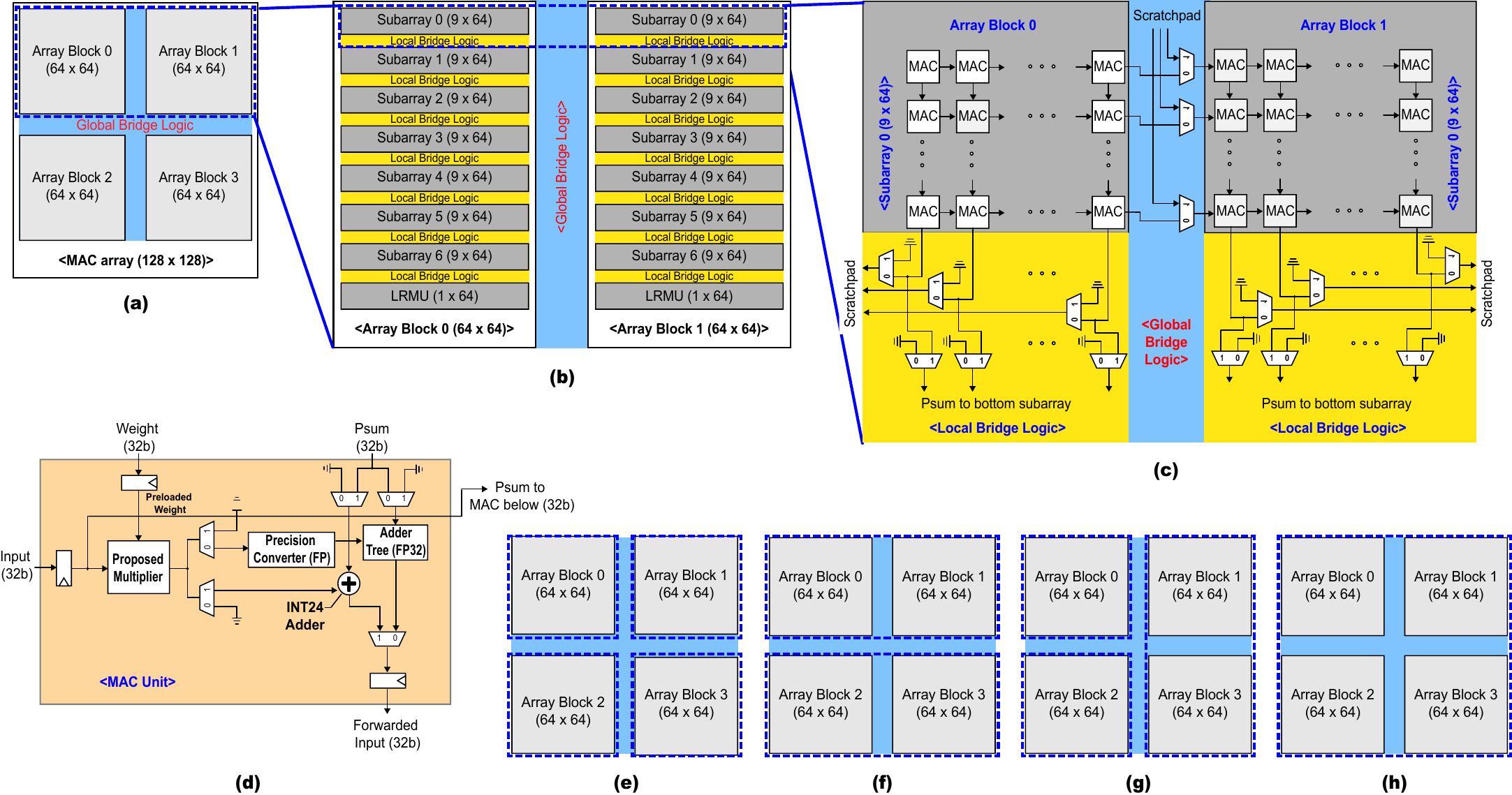}    
    \caption{(a) Overall structure of fusible and fissionable MAC array. (b) Structure of array blocks. (c) Structure of subarrays and bridge logics. (d) Schematic diagram of MAC unit. (e-h) Examples of fused array blocks: (e) four 64$\times$64 MAC arrays, (f) two 64$\times$128 MAC arrays, (g) two 64$\times$64 MAC arrays and one 128$\times$64, and (h) one 128$\times$128 MAC array.}
    \label{fig:mac_array_overall}
\end{figure*}

\textbf{Advantage of the proposed multiplier:} The proposed multiplier enables support for AI models with various data types while maintaining low hardware cost.
For example, recent deep learning training process exploits bfloat16~\cite{nnp-t, bfloat16} or FP8~\cite{parameter8_training, fp8_training_2022, hybrid_fp8_training}.
Additionally, INT4 and INT8 are widely employed in state-of-the-art post training quantization (PTQ)~\cite{ptq_4bit, ptq_eccv_8bit, ptq_calibration} and inference processes~\cite{int4_nvidia, inference_8bit_transformer}. 
Our all-in-one multiplier can efficiently support these AI models across various data formats.
Another notable point is that the hardware design does not require separate multipliers to manage scaling factors.
Many recent studies on the training of low-bit quantized models frequently utilize exponential scaling factors to compensate for the limited dynamic range of quantized data~\cite{parameter8_training, fp8_training_2022, nvidia_mixed_precision, hybrid_fp8_training}. 
Since the proposed multiplier can accept programmable bias values, it eliminates the need for additional multipliers to incorporate these scaling factors.

\section{Flexibly Fusible and Fissionable MAC Array}\label{sec:structure_mac_array}

The TPU-like SA structure provides high data reusability through its regular architecture and simple control mechanisms. 
Thus, this structure has been widely adopted in various industrial and academic AI accelerators~\cite{planaria, neummu, TPU_v4, data_flow_mirror, sigma, sara, bitfusion}. 
However, the traditional SA structure achieves high MAC utilization only for specific AI operations, data dimensions, and the execution of a single AI model. 
To overcome the limitation of the conventional SA structure, we propose novel mapping schemes tailored to different AI operations, along with a MAC array structure designed to support them.

\subsection{Microarchitecture of a Morphable MAC Array}\label{sec:overall_structure_mac_array}
Fig.~\ref{fig:mac_array_overall}-(a-d) show the design hierarchy of our proposed morphable MAC array. 
The MAC array consists of four array blocks, each containing seven subarrays and one Last Row MAC Unit (LRMU).
Each subarray comprises 9$\times$64 MAC units, and the LRMU consists of 1$\times$64 MAC units. 
Multiplexers are positioned between subarrays and array blocks as shown in Fig.~\ref{fig:mac_array_overall}-(c). 
These multiplexers serve as data routers by connecting subarrays and an LRMU within the array block or fusing subarrays and LRMUs between the array blocks. 
The multiplexer logic in the former case is referred to as local bridge logic, while in the latter, it is referred to as global bridge logic. 
Fig.~\ref{fig:mac_array_overall}-(d) presents a block diagram of a MAC unit designed with our proposed multiplier.

\begin{figure*}[t]
    \centering
    \includegraphics[scale=0.50]{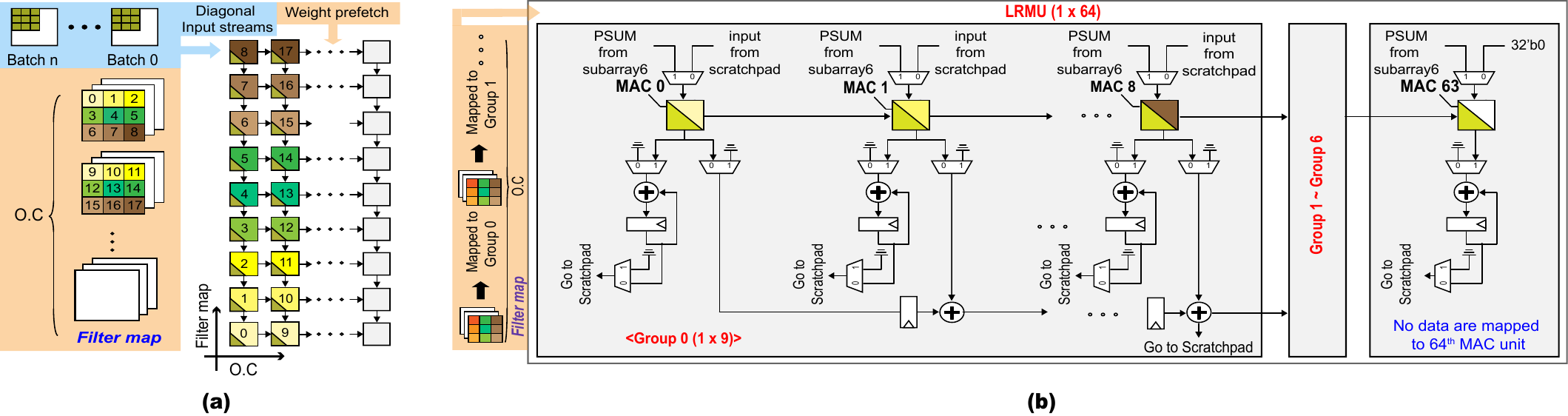}
    \caption{Data mapping on (a) one subarray and (b) one last row MAC unit (LRMU) in the AI operation in which MAC results are not accumulated in the direction of the input channel. }
    \label{fig:lrmu}    
\end{figure*}

\subsection{Data Mapping with Our Morphable Structure}
As presented in Table~\ref{tab:classification_of_operations}, AI operations can be categorized into two types: those where the results of MAC operations are accumulated in the direction of the input channel and those where they are not. 
Since the proposed MAC array employs distinct mapping schemes based on the type of operations, we examine how data are mapped to a single array block of the proposed MAC array for each operation type, and how the sub-modules of the array block are reconfigured to accommodate the different data mapping schemes.

\textbf{Accumulable operations:} In the operations in which MAC results are accumulated along the input channel, local bridge logics connect the subarrays and the LRMU within an array block, allowing the array block to work as a 64$\times$64 TPU-like SA structure. 
Specifically, for input channel-wise accumulable convolution operations, the filter data corresponding to the '$C_{in}$' and '$C_{out}$' dimensions are loaded into the columns and rows of the array block, respectively, prior to streaming the input data (Fig.~\ref{fig:motivation_a}-(a)). 
In the case of GEMM operations, a 64$\times$64 weight matrix is preloaded into the array block. 
Once the prefetching is complete, the corresponding input feature map or input matrix data are streamed diagonally into the array block, and computation is started.

\textbf{Unaccumulable operation:} In the operations in which MAC results are not accumulated along the input channels, local bridge logics do not connect subarrays and the LRMU within an array block. 
This means that partial sums generated in each array block are not delivered to other subarrays or the LRMU. 
Fig.~\ref{fig:lrmu}-(a) illustrates how data are mapped to each subarray within an array block for these operations.
Elements of filters are mapped to the columns of each subarray, and elements of the '$C_{out}$' dimension are mapped to the rows.
In the case of LRMU, 9 MAC units form one group, and elements of one filter are mapped to the one group as shown in Fig.~\ref{fig:lrmu}-(b). 
Since there are 64 MAC units in LRMU, 7 groups are created. 
Therefore, out of the 64 MACs in the LRMU, 7 (\# of groups) $\times$ 9 (\# of MACs within a group) = 63 MAC units are utilized for computations.
As a result, the array block can provide $>$99\% MAC utilization even in the non-accumulating operations.
To ensure seamless synchronization in this operation mode, the output data from the LRMU is transferred to the scratchpad memory through a pipelined datapath implemented using flip-flops.

\subsection{Fusible and Fissionable MAC Array}\label{sec:fusible_mac_array}

The proposed MAC array can be reconfigured into various sizes through the global bridge logics between array blocks, as shown in Fig.~\ref{fig:mac_array_overall}-(e-h). 
For example, the MAC array can be partitioned into four 64$\times$64 MAC arrays by disconnecting all global bridge logics (Fig.~\ref{fig:mac_array_overall}-(e)), or into two 64$\times$128 arrays by connecting the bridge logics between array blocks 0 and 1, as well as between array blocks 2 and 3 (Fig.~\ref{fig:mac_array_overall}-(f)). 
Additionally, the MAC array can be configured to support one 128$\times$64 array alongside two 64$\times$64 arrays (Fig.~\ref{fig:mac_array_overall}-(g)), or it can form a single 128$\times$128 MAC array by connecting all global bridge logics between array blocks, as shown in Fig.~\ref{fig:mac_array_overall}-(h). 
This reconfigurability enables the efficient execution of GEMM operations across various sizes and is particularly advantageous for processing multi-tenant execution of multiple AI models within a single MAC array. 
For instance, four AI models can be processed in parallel when the MAC array is configured into four 64$\times$64 MAC arrays, as depicted in Fig.~\ref{fig:mac_array_overall}-(e), and two AI models can be processed simultaneously using two 64$\times$128 arrays, as shown in Fig.~\ref{fig:mac_array_overall}-(f).

\begin{figure}[t]
    \centering
    \includegraphics[scale=0.60]{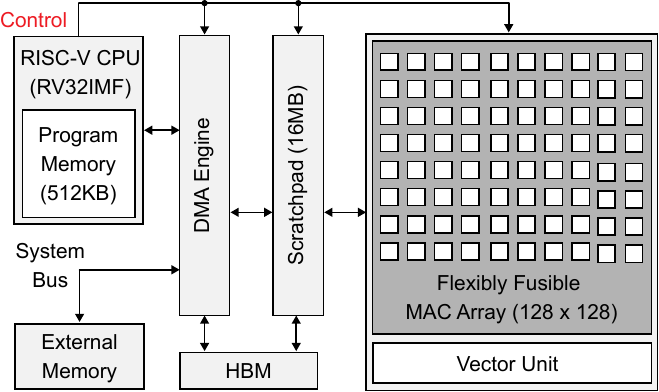}
    \caption{Overall architecture of the proposed All-rounder accelerator.}
    \label{fig:overall_structure}
\end{figure}

\section{All-rounder Architecture and Customized ISA}

\subsection{Overall Architecture of All-rounder Accelerator}\label{sec:structure_all_round_tpu}
Fig.~\ref{fig:overall_structure} illustrates the overall  microarchitecture of All-rounder accelerator. 
This accelerator comprises a flexible MAC array, a vector unit, 16MB scratchpad memory (SPM), a direct memory access (DMA) engine, high bandwidth memory (HBM) and a RISC-V CPU. 
The MAC array consists of 128$\times$128 MAC units, \textit{each designed with the proposed all-in-one multiplier}. 
The MAC array ensures high MAC utilization by fusing or splitting its structure based on AI operation types or multi-tenant scenarios. 
The vector unit, consisting of 128 ALUs, handles operations for non-DNN layers, such as ReLU/GELU, batch (or layer) normalization, and softmax functions, including data quantization.
The MAC array and the vector unit fetch/store data from/to 16MB SPM\footnote{We opted for an SPM instead of a hardware-managed SRAM cache due to its lack of tags and associative sets, which makes it more area-efficient~\cite{buffets, stash}. In addition, the SPM compiler easily predicts memory access patterns owing to deterministic dataflow of AI operations~\cite{neummu}.}. 
Since the SPM is double buffered in our system, 8MB of SPM is used for data prefetching from the external host memory through the DMA engine, while the remaining capacity feeds data to the MAC array and vector unit or stores partial sums generated by the array or a subset of units. 
Furthermore, the All-rounder accelerator can directly load both programs and data from the host memory through the DMA engine.
Thanks to the inherent 5-stage pipelined CPU of All-rounder, programs can be decoded within the accelerator. 
The CPU supports the 32-bit RISC-V base instruction set along with extended instruction sets of M type (for multiplication and division) and F type (for single-precision FP operations)~\cite{risc_v_manual}.
Additionally, the CPU supports instructions of interrupt, halt, synchronization, and no operation (NOP).

\begin{figure}[t]
    \centering
    \includegraphics[scale=0.68]{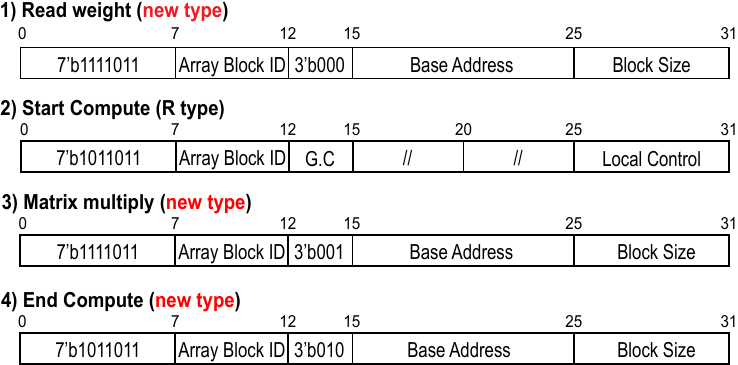}
    \caption{Instructions utilized in the execution of the morphable MAC array.}
    \label{fig:instruction}
\end{figure}

\subsection{Customized Instructions for Morphable MAC Array}\label{sec:Customized_ISA}

The All-rounder accelerator leverages customized instruction sets to execute AI operations within the MAC array instead of the repetitive use of vector instructions of RISC-V ISA. 
The opcodes of the instructions are defined with values of 7'b1011011 and 7'b1111011, which are guided values for the custom field outlined in~\cite{risc_v_manual}, and a new instruction format is defined to support the customized instructions. 
Fig.~\ref{fig:instruction} depicts the instructions utilized in our proposed MAC array. 
The instructions are defined using the `R type' of RISC-V ISA and a newly defined format.
In the All-rounder architecture, the following instructions are executed in sequence: i) \texttt{Read weights}, ii) \texttt{Start compute}, iii) \texttt{Matrix multiply}, and iv) \texttt{End compute}.
The \texttt{Read weights} instruction is executed first, loading weight data of size `64$\times$\{variable block size\}' from the base address of the SPM to the array block corresponding to the ID.
Once prefetching is complete, \texttt{Start compute} sets control signals of the array block. 
The control signals are determined by the `func3' and `func7' fields of the instruction. 
The `func3' field (G.C in Fig.~\ref{fig:instruction}) generates global control signals that determine whether to fuse or split the array blocks, while the `func7' produces local control signals for the interior of the array block based on the information of operation mode, precision, and data type.
\texttt{Matrix multiply} is then decoded and transferred to the array block after the control signals have been set. 
When this instruction is executed, input data is fetched with the size of `64$\times$\{variable block size\}' from the base address of SPM to the corresponding array block, and the MAC array begins its computation. 
Finally, \texttt{End compute} writes outputs of the array block to the SPM and refreshes the array block so that it can fetch new weight data.

\section{Evaluation}
We designed our proposed multiplier and all building blocks of the All-rounder accelerator using Verilog. 
To analyze the benefit of using our all-in-one multiplier, we implemented two multi-format multiply units as baselines for the comparison.
To evaluate the practical effectiveness of All-rounder accelerator, we also implemented three representative AI accelerators.
Moreover, we assessed the performance and energy efficiency of the All-rounder accelerator against a high-end GPU device. 
This section presents the experimental setup and results.

\subsection{Delay/Area/Energy analysis of the Proposed Multiplier}\label{sec:analysis_multiplier} 
\textbf{Baselines and methodology:} State-of-the-art quantization techniques exploit formats such as bfloat16~\cite{bfloat16}, FP8 (i.e., FP8-A\{s:1, e:4, m:3\} or FP8-B\{s:1, e:5, m:2\})~\cite{parameter8_training, fp8_training_2022, hybrid_fp8_training, hybrid_fp8_a}, INT8~\cite{ptq_eccv_8bit, inference_8bit_transformer}, and INT4~\cite{ptq_4bit, int4_nvidia}. 
Accordingly, we implemented two multiply units for comparison that support these formats with same throughput as our all-in-one multiplier.
These baselines integrate dedicated multipliers for each supported format, similar to off-the-shelf accelerators~\cite{nvidia_h100, TPU_v4, edge_rapid, gpnpu}. 
Specifically, one of the baselines consists of a bfloat16 multiplier, four FP9 multipliers that support both FP8 formats (i.e., FP8-A and FP8-B) and a multiply unit akin to Bit Fusion~\cite{bitfusion} that supports INT4 and INT8 formats. 
The other mirrors the structure of IBM RaPiD~\cite{edge_rapid} but is configured with different precisions, comprising one bfloat16 multiplier, four FP9 multipliers, one INT8 multiplier, and four INT4 multipliers.
To evaluate hardware efficiency, we synthesized the all-in-one multiplier and two baseline designs using Synopsys Design Compiler (ver. N-2017.09-SP5) with a 28nm CMOS library. 
We report the maximum achievable operating frequencies for these designs without causing setup time violations. 
To ensure fairness in comparisons, we also provide the area and energy per operation (energy/op) of the all-in-one multiplier and the two baseline designs, synthesized at the same target frequency (i.e., $T_{clk}$ = 2.5ns). 
For the energy analysis, power metrics were extracted using Synopsys PrimePower (ver. P-2019.03-SP5) with the gate-level netlist and Synopsys Design Constraint (SDC) file.

\begin{table*}[t]
    \centering
    \caption{Comparison between the proposed all-in-one multiplier and two baselines} 
    \includegraphics[scale=0.27]{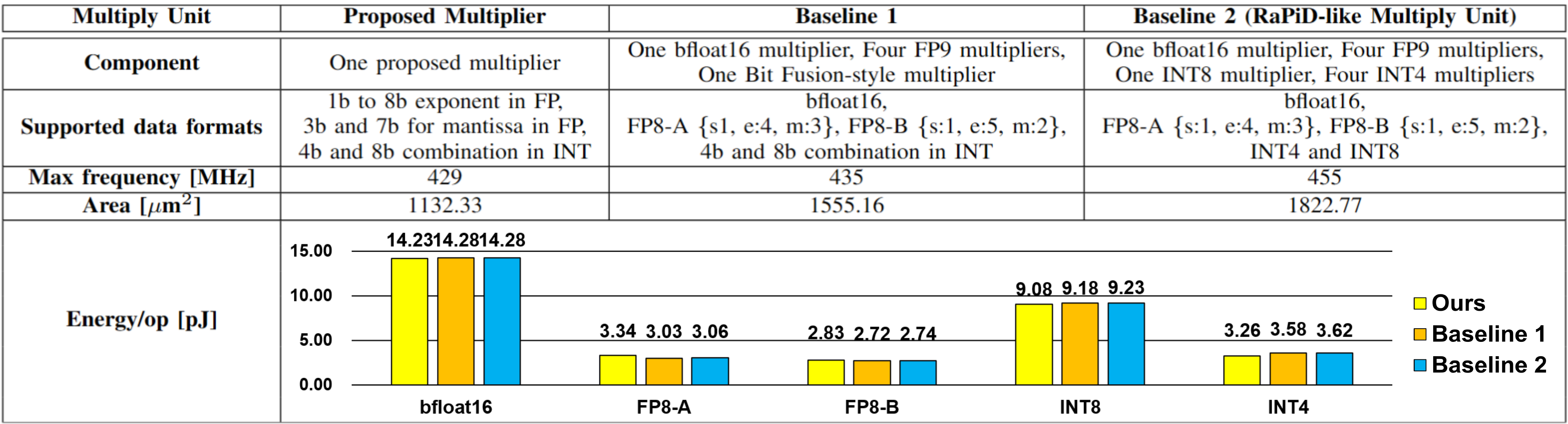}
    \label{tab:multiply_unit_analysis}
\end{table*}

\textbf{Analysis with experimental results: }Table~\ref{tab:multiply_unit_analysis} summarizes the design configurations of the proposed multiplier and baseline multiply units, along with their area and energy values.
The proposed multiplier supports exponents ranging from 1b to 8b and mantissas with multiple precisions of 3b and 7b, allowing it to handle a wider range of precisions compared to the baselines in FP operation modes.
Although the proposed multiplier operates at a lower frequency of 429 MHz (i.e., $T_{clk}$ = 2.33ns), which is 1.38\% and 5.71\% lower than the baseline designs due to the increased critical path caused by the reconfigured CSM, it achieves notable area savings.
Specifically, the all-in-one multiplier occupies only 1132.33 µm\textsuperscript{2}, which is 1.37$\times$ and 1.61$\times$ smaller than the baselines. 
This reduction is primarily due to the proposed multiplier having only a single CSM, which occupies the largest area among the sub-modules of the FP multiplier (Fig.~\ref{fig:fp_multiplier}), unlike the baseline multiply units.
In terms of energy efficiency, the proposed multiplier achieves energy savings of 0.37$\sim$9.31\% in bfloat16, INT8, and INT4 modes compared to the baselines. 
However, it consumes 6.11\% more energy on average in FP8 modes due to the activated isolation logic, which gates the sub-modules (e.g., normalizers and rounders depicted in Fig.~\ref{fig:multiplier_examples}-(a)) in the CSM and the normalizer when they are not in use.

\begin{figure}[t]
    \centering
    \includegraphics[scale=0.62]{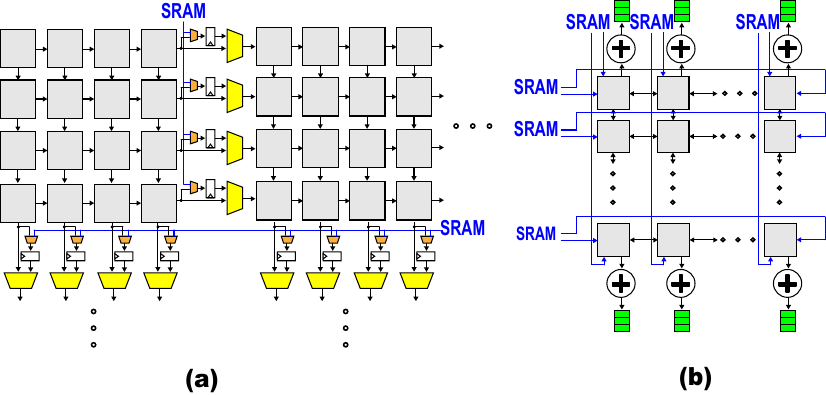}
    \caption{Baseline MAC array structures: (a) and (b) depict the architectures of~\cite{sara} and~\cite{data_flow_mirror}, respectively.}
\label{fig:baseline_architectures} 
\end{figure}

\subsection{Architectural Analysis with Single-Tenant Execution}\label{sec:standalone_analysis}

\textbf{Baselines and methodology: }To evaluate the benefits of the All-rounder accelerator, we designed a rigid systolic array (SA) (i.e., Google TPU~\cite{TPU_v4}) as shown in Fig.~\ref{fig:motivation_a}. 
Additionally, we implemented two state-of-the-art MAC arrays from~\cite{data_flow_mirror} and~\cite{sara}, which modify the rigid SA in order to support flexibility within the SA structure. 
Fig.~\ref{fig:baseline_architectures} illustrates the architectures of these flexible MAC arrays. 
The MAC array of~\cite{sara}, depicted in Fig.~\ref{fig:baseline_architectures}-(a), divides the TPU-like SA into 4$\times$4 systolic cells and adds bypass links (i.e., multiplexers) at the cell edges. 
This design allows the MAC array to support various GEMM dimensions by connecting or disconnecting the cells via the bypass links. 
The MAC array of~\cite{data_flow_mirror}, shown in Fig.~\ref{fig:baseline_architectures}-(b), replaces the omni-directional communication network among the MAC units in the TPU-like SA with a bi-directional communication network. 
Furthermore, this MAC array includes additional hardware components to enable data mirroring for bidirectional streaming of input data and bidirectional collection of partial sums.
To conduct an architectural comparison, we synthesized three baselines, i.e., the TPU-like SA and the MAC arrays of~\cite{sara} and~\cite{data_flow_mirror}, equipped with the proposed all-in-one multipliers.
We also considered SRAM and DRAM accesses when estimating performance and energy consumption.
We assumed that the baselines had the same memory configuration as the All-rounder accelerator shown in Fig.~\ref{fig:overall_structure}.
The energy consumption of the SPM was modeled using CACTI-P~\cite{cactip}, and the power and timing specifications of DRAM were based on~\cite{hbm2}. 
Finally, we estimated MAC utilization and clock cycles by modifying an open-source cycle-level simulator, SCALE-sim~\cite{scale_sim}, to reflect our SRAM and DRAM configurations.
To evaluate the latency of computing the product of the input matrix (dimensions $\{S_C, T\}$) and the weight matrix (dimensions $\{T, S_R\}$) within the MAC arrays, we used the following formula~\footnote{When calculating the latency of CNN models, we followed the approach described in~\cite{scale_sim}, converting the input feature maps and filters into the input matrix and weight matrix, respectively. 
For unaccumulable CNN models, as discussed in Section~\ref{sec:motivation_a_content}, the values of $R$ and $C$ in Eq. (\ref{eq:rn_bw_eq0}) for the baseline MAC arrays were determined by the output bus bandwidth constraint.
}:
\begin{equation}\label{eq:rn_bw_eq0}
\text{Latency} = (2S_R + S_C - 2) \times \left(\left\lceil \frac{S_R}{R} \right\rceil \times \left\lceil \frac{S_C}{C} \right\rceil \right),
\end{equation}
where \(R\) and \(C\) represent row and column dimensions of the MAC array. 
For non-morphable architectures (e.g., TPU-like SA and the MAC array in~\cite{data_flow_mirror}), we used fixed values of \(R = C = 128\) in both bfloat16 and INT8 modes, and \(R = C = 256\) in FP8 and INT4 modes. 
In contrast, for the morphable MAC arrays (e.g., All-rounder and the MAC array in~\cite{sara}), we configured \(R, C \in \{64, 128\}\) in bfloat16 and INT8 modes, and \(R, C \in \{128, 256\}\) was used in FP8 and INT4 modes.

\textbf{AI benchmarks:}
The hardware evaluation was conducted on five real-world CNN models: VGG16~\cite{vggnet}, ResNet-18~\cite{resnet}, MobileNetV2~\cite{mobilenet_v2}, EfficientNet-B0~\cite{effcientnetv1}, and ConvNeXt-S~\cite{convnext}, as well as two representative large language models (LLMs): GPT-2 small~\cite{gpt2} and Llama-2 (7B)~\cite{llama2}.
For the CNN models, we used ImageNet dataset~\cite{imagenet}, and the LLMs were evaluated on WikiText-2 dataset~\cite{wikitext}.
We selected prior studies in~\cite{bfloat16} and~\cite{hybrid_fp8_training}, which utilize the bfloat16 and hybrid FP8 formats (i.e., FP8-A and FP8-B), respectively, to analyze the benchmarks during the training process. 
The evaluation was performed over a single training iteration with a batch size ($B$) of 128 for CNN models and 8 for LLMs.
Also, we report the evaluation results for the inference process using integer formats (i.e., INT8 and INT4) based on previous works~\cite{int4_nvidia, inference_8bit_transformer}.

\begin{table*}[t]
\centering
\caption{Supported hardware configurations of the All-rounder accelerator and its area/power comparison to the baseline accelerators (i.e., conventional SA, SA with SARA~\cite{sara} and dataflow mirroring~\cite{data_flow_mirror})}
\label{tab:design_specifications}
\scalebox{0.93}{%
\begin{tabular}{|c|c|c|c|c|}
\hline
\textbf{Hardware}                               & \textbf{Proposed MAC array}                                                                                         & \textbf{TPU-like SA}                                                                                               & \textbf{~\cite{sara}-based accelerator}                                                                                          & \textbf{~\cite{data_flow_mirror}-based accelerator}                                                                                         \\ \hline\hline
\textbf{Technology {[}nm{]}}           & 28                                                                                                                  & 28                                                                                                                 & 28                                                                                                                  & 28                                                                                                                 \\ \hline
\textbf{Operating frequency {[}MHz{]}} & 400                                                                                                                 & 400                                                                                                                & 400                                                                                                                 & 400                                                                                                                \\ \hline
\textbf{Array size} & 128$\times$128                                                                                                                 & 128$\times$128                                                                                                                & 128$\times$128                                                                                                                 & 128$\times$128                                                                                                                \\ \hline
\textbf{\# of Multipliers}                    & \begin{tabular}[c]{@{}c@{}}128$\times$128 (bfloat16 or INT8)\\ 256$\times$256 (FP8 or INT4)\end{tabular}                            & \begin{tabular}[c]{@{}c@{}}128$\times$128 (bfloat16 or INT8)\\ 256$\times$256 (FP8 or INT4)\end{tabular}                           & \begin{tabular}[c]{@{}c@{}}128$\times$128 (bfloat16 or INT8)\\ 256$\times$256 (FP8 or INT4)\end{tabular}                            & \begin{tabular}[c]{@{}c@{}}128$\times$128 (bfloat16 or INT8)\\ 256$\times$256 (FP8 or INT4)\end{tabular}                           \\ \hline
\textbf{Area {[}mm\textsuperscript{2}{]}}                 & 108.03                                                                                                              & 103.55                                                                                                             & 118.45                                                                                                              & 105.84                                                                                                             \\ \hline
\textbf{Power {[}W{]}}                 & \begin{tabular}[c]{@{}c@{}}5.31 (bfloat16) \\ 10.14 (FP8-A)\\ 9.19 (FP8-B)\\ 1.73 (INT8)\\ 1.70 (INT4)\end{tabular} & \begin{tabular}[c]{@{}c@{}}4.73 (bfloat16) \\ 9.57 (FP8-A)\\ 8.62 (FP8-B)\\ 1.16 (INT8)\\ 1.14 (INT4)\end{tabular} & \begin{tabular}[c]{@{}c@{}}6.32 (bfloat16) \\ 11.16 (FP8-A)\\ 10.21 (FP8-B)\\ 2.75 (INT8)\\ 2.73 (INT4)\end{tabular} & \begin{tabular}[c]{@{}c@{}}4.92 (bfloat16) \\ 9.74 (FP8-A)\\ 8.77 (FP8-B)\\ 1.30 (INT8)\\ 1.28 (INT4)\end{tabular} \\ \hline
\end{tabular}}
\end{table*}

\begin{figure}[t]
    \centering
    \includegraphics[scale=0.60]{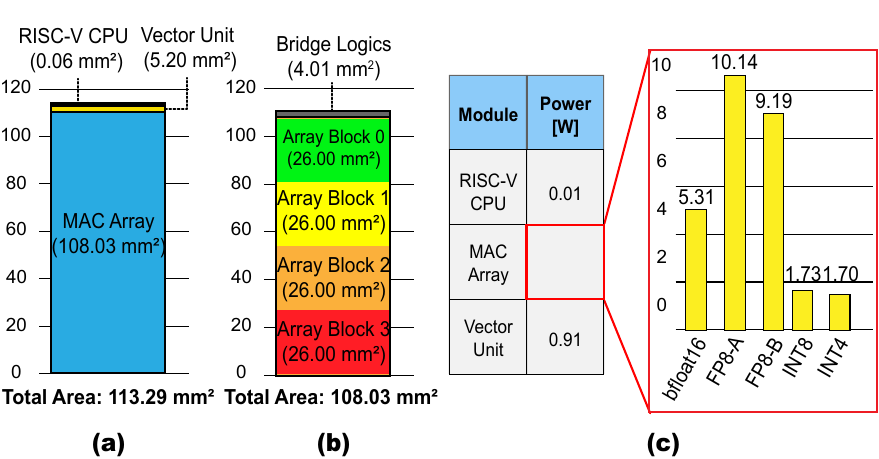}
    \caption{Area breakdowns of (a) All-rounder accelerator and (b) the proposed morphable MAC array. (c) Power breakdown of the All-rounder accelerator.}
\label{fig:area_power_breakdown_all_rounder}    
\vspace{-2mm}
\end{figure}

\textbf{Area and power analysis:} Fig.~\ref{fig:area_power_breakdown_all_rounder}-(a) and (b) present the area breakdowns of the All-rounder accelerator and the morphable MAC array, respectively.
The total area of the accelerator is 113.29 mm$^2$, and the proposed MAC array occupies 108.03mm$^2$. 
The bridge logic circuits, which facilitate the connection or separation of subarrays or array blocks, account for 3.71\% (i.e., 4.01 mm$^2$) of the total area of the MAC array.
Fig.~\ref{fig:area_power_breakdown_all_rounder}-(c) illustrates power breakdown of the All-rounder accelerator. 
Additionally, it reports the power consumed by the MAC array across different data formats (represented by yellow bars).
The proposed MAC array consumes 1.70$\sim$10.14W of power, and the bridge logics account for 10.73\% of the total power on average across data formats.

\begin{figure*}[t]
    \centering
    \includegraphics[scale=0.75]{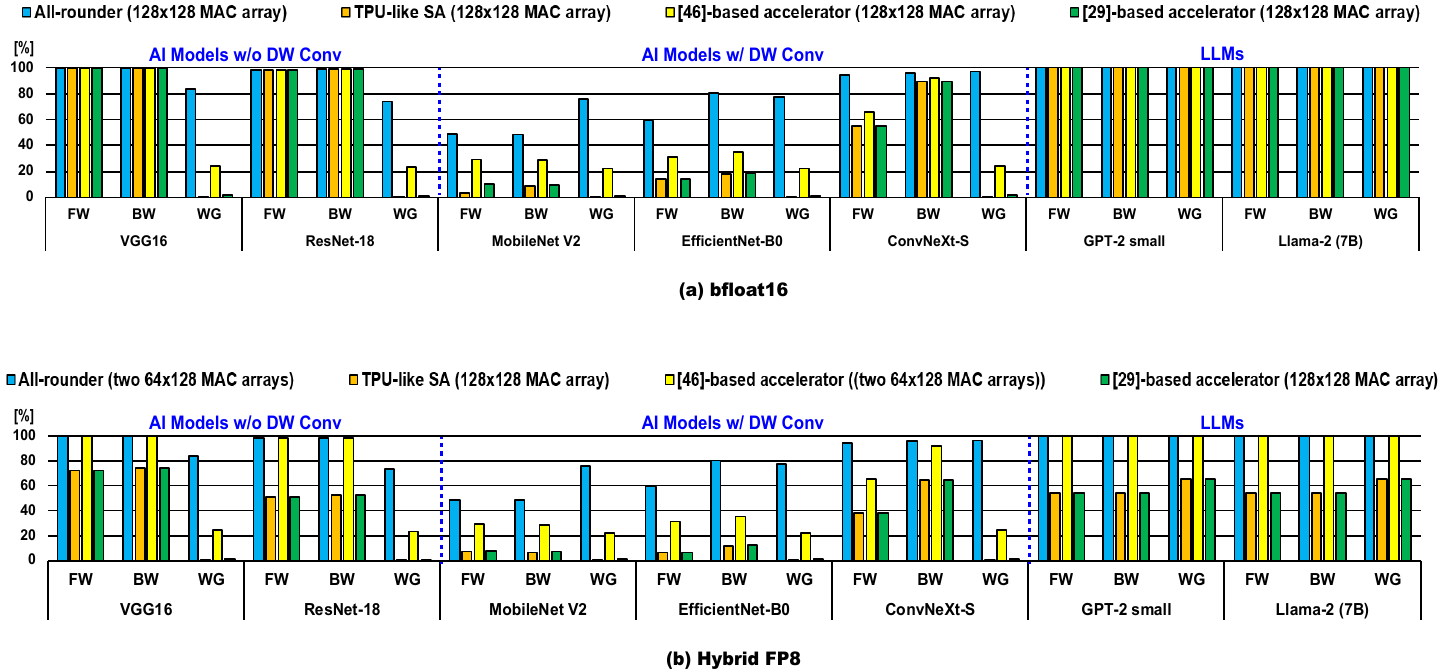}
    \caption{MAC utilization during the training process in (a) bfloat16 and (b) hybrid FP8 formats. In the figure, `FW', `BW,' and `WG' denote the forward pass, backward pass, and weight gradient step, respectively. Bars in sky blue, orange, yellow, and green represent the MAC utilization of the All-rounder accelerator, TPU-like systolic array, the accelerator based on~\cite{sara} and the accelerator based on~\cite{data_flow_mirror}, respectively.}    
    \label{fig:utilization_result}
\end{figure*}

\textbf{Analysis with design specification: }Table~\ref{tab:design_specifications} summarizes the design specifications of our MAC array and the baseline accelerators. 
Since critical paths of all designs occur in the multiplier, all the accelerators have the same operating frequency of 400MHz. 
Although all accelerators consist of 128$\times$128 MAC units, the number of multipliers varies depending on the data formats. 
For instance, the designs feature 128$\times$128 multipliers for bfloat16 and INT8 formats, while they include 512$\times$512 multipliers for FP8 and INT4 formats. 
This is because, as illustrated in Fig~\ref{fig:multiplier_examples}, the all-in-one multiplier produces one multiplication result for bfloat16 and INT8 formats, but generates four multiplication results for FP8 and INT4 formats. 
Table~\ref{tab:design_specifications} also compares the area and power consumption. 
The proposed array structure requires 4.33\% and 2.07\% larger areas than the TPU-like SA and the~\cite{data_flow_mirror}-based accelerator, respectively, yet it occupies 8.80\% smaller area than the \cite{sara}-based accelerator.
In addition, the proposed MAC array consumes 1.25$\times$ and 1.15$\times$ more power on average than the TPU-like SA and the \cite{data_flow_mirror}-based accelerator, respectively, while consuming 1.32$\times$ less power on average than the \cite{sara}-based accelerator.
These differences in area and power consumption stem from logic structures designed to support flexibility.

\textbf{Comparison of MAC utilization: }We report the MAC utilization when data is mapped onto a 128$\times$128 MAC array in both bfloat16 and INT8 modes.
In hybrid FP8 and INT4 modes, we provide the MAC utilization for morphable MAC arrays (i.e., All-rounder and~\cite{sara}-based accelerator) when data is mapped onto two 64$\times$128 MAC arrays.
For non-morphable MAC arrays (i.e., TPU-like SA and~\cite{data_flow_mirror}-based accelerator), data is mapped onto a 128$\times$128 MAC array.
Fig.~\ref{fig:utilization_result}-(a) illustrates MAC utilization during the training process in bfloat16.
When evaluating CNNs without depthwise convolutions, such as VGG16 and ResNet-18, the~\cite{data_flow_mirror}-based accelerator and the TPU-like SA show low MAC utilization only during the weight gradient (WG) step of the training process. 
This inefficiency arises from bottlenecks stemming from the narrow output buses of MAC arrays connected to accumulators.
Both the~\cite{sara}-based accelerator and the All-rounder demonstrate improved utilization; however, our accelerator achieves 3.29$\times$ higher utilization on average during the WG step compared to the~\cite{sara}-based accelerator.
For models including depthwise convolutions, such as MobileNetV2, EfficientNet-B0, and ConvNeXt-S, the All-rounder accelerator achieves superior utilization across all training steps compared to other baseline accelerators.
This improvement comes from the ability of the All-rounder to employ distinct data mapping schemes tailored to AI operation types, supported by its configurable architecture.

For LLMs, we utilized a sequence length of $L=512$, a model dimension ($d_{\text{model}}$) of 4,096, and a head dimension ($d_{\text{head}}$) of 128.
Then, key, query, and value matrices for each attention head
will have dimension of $\mathbb{R}^{4,096 \times 128}$, where $B\cdot L=4,096$.
Since dimensions of the per-head key, query, and value matrices are multiples of 128, the resource utilization of the LLMs approaches nearly 100\% for all accelerators in bfloat16.
Meanwhile, Fig.~\ref{fig:utilization_result}-(b) presents MAC utilization during training with the hybrid FP8 format. 
In the non-morphable accelerators, i.e., TPU-like SA and~\cite{data_flow_mirror} and the one described in~\cite{data_flow_mirror}, the data dimensions are smaller than the number of multipliers, resulting in reduced MAC utilization compared to the bfloat16 format.
Nevertheless, the accelerator described in~\cite{sara} universally achieves higher MAC utilization than these accelerators, owing to its morphable architecture.
Moreover, the All-rounder accelerator attains a 1.31$\times$ higher MAC utilization on average than the accelerator presented in~\cite{sara} when evaluated on the target AI benchmarks. 
This improvement stems from the distinct data mapping scheme employed by the All-rounder accelerator.
Although not depicted in the figure, the All-rounder achieves enhanced MAC utilization compared to the TPU-like SA,~\cite{sara}-based accelerator, and
~\cite{data_flow_mirror}-based accelerator, with improvements of 3.96$\times$, 1.29$\times$, and 2.09$\times$, respectively, for INT8 inference, and 3.39$\times$, 1.25$\times$, and 3.21$\times$, respectively, for INT4 inference.

\begin{figure*}[ht]
    \centering
    \includegraphics[scale=0.73]{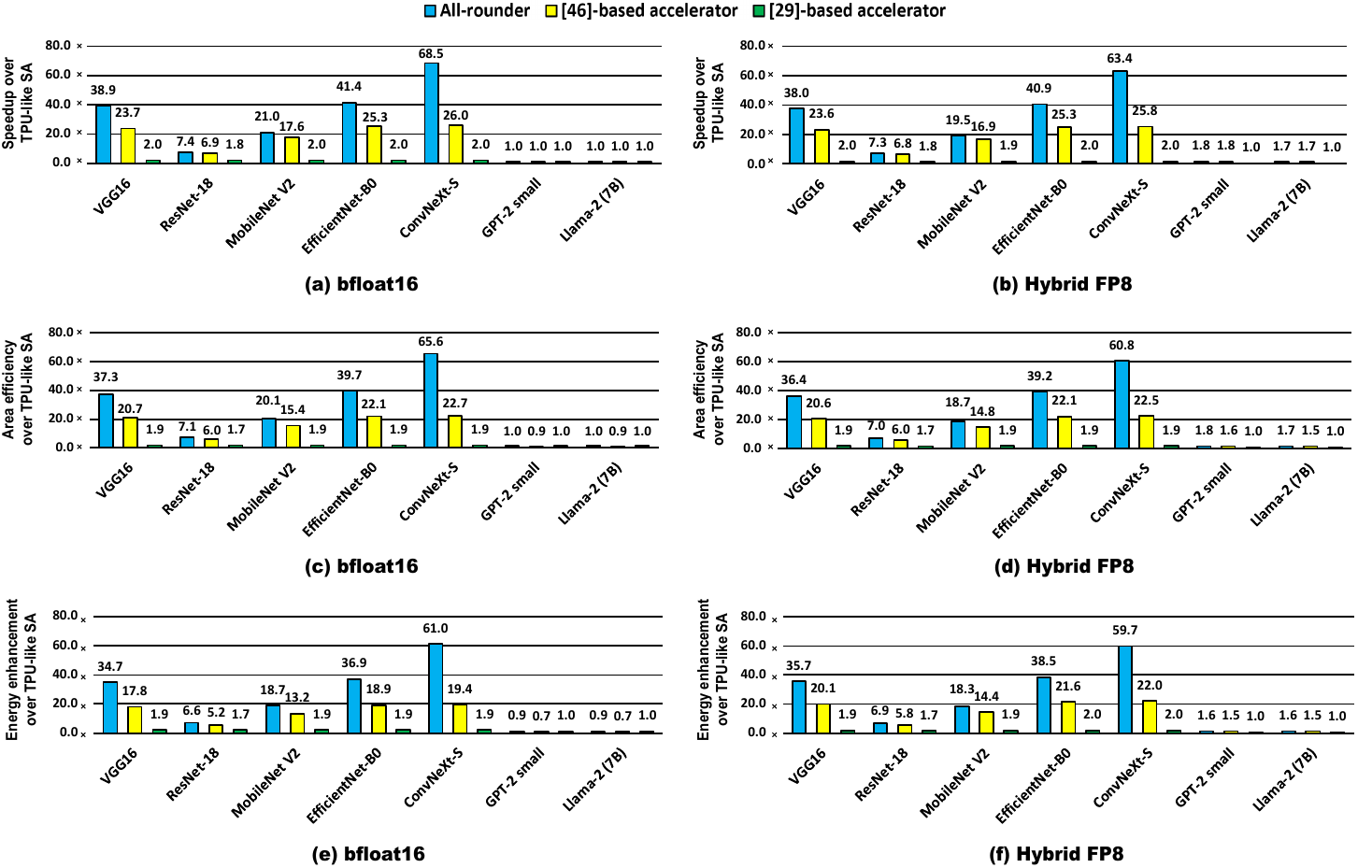}
    \caption{
    (a-b) Performance, (c-d) area efficiency (defined as throughput per MAC array area), and (e-f) energy efficiency improvements of All-rounder and two baseline accelerators (i.e., the accelerators based on~\cite{sara} and~\cite{data_flow_mirror}) compared to the TPU-like systolic array during the training process with bfloat16 and hybrid FP8 formats.}  
    \label{fig:evaluation_result}
\end{figure*}

\textbf{Hardware efficiency improvements: }Fig.\ref{fig:evaluation_result}-(a) and (b) show the performance benefits of the All-rounder and baseline accelerators over the typical TPU-like SA. 
The All-rounder accelerator achieves an average speedup of 25.60$\times$ in bfloat16 and 24.65$\times$ in the hybrid FP8 format. 
This represents 1.77$\times$ and 15.33$\times$ improvements over the~\cite{sara} and~\cite{data_flow_mirror} accelerators in bfloat16, and 1.87$\times$ and 14.90$\times$ improvements over the~\cite{sara} and~\cite{data_flow_mirror}-based accelerators in the FP8 format. 
Fig.~\ref{fig:evaluation_result}-(c) and (d) illustrate the area efficiency improvements with the TPU-like SA as the baseline.
The All-rounder achieves 1.94$\times$ and 1.84$\times$ higher area efficiency, on average, over the~\cite{sara}-based accelerator in the bfloat16 and hybrid FP8 formats, respectively.
It also demonstrates area efficiency improvements of 15.02$\times$ and 14.48$\times$ on average compared to the~\cite{data_flow_mirror}-based accelerator accordingly.
In addition, Fig.~\ref{fig:evaluation_result}-(e) and (f) show energy efficiency improvements using the TPU-like SA as the baseline.
The All-rounder achieves an average improvement of 22.82$\times$ in the bfloat16 and 23.20$\times$ in the hybrid FP8 format.
These improvements in performance, area, and energy efficiency are attributed to the high MAC utilization of the proposed accelerator, regardless of the data type of AI operations.
In INT8 and INT4 modes, which are not included in Fig.~\ref{fig:evaluation_result}, the All-rounder reduces the energy consumption by 74.48\% and 90.97\%, respectively, compared to the TPU-like SA, and achieves 1.83$\times$ and 2.24$\times$ higher energy efficiency, on average, than \cite{sara}-based and \cite{data_flow_mirror}-based accelerators in INT8, and 2.54$\times$ and 6.56$\times$ higher energy efficiency in INT4.

\vspace{-2mm}
\subsection{Architectural Evaluation for Multi-Tenant Execution}\label{sec:multitenant_analysis}

\textbf{Baselines and methodology: }
For a more realistic analysis, we evaluated the runtime in a multi-tenant scenario. 
We selected AI models from two online applications: image captioning and image classification. 
The models for image captioning comprise a CNN model that extracts image features and an NLP model that maps these features into words~\cite{image_captioning, dnpu}. 
The NLP model is connected to the end of the CNN model, and we assumed that all models are quantized to INT8 based on~\cite{inference_8bit_transformer}. 
To analyze the multi-tenant scenario, we evaluated the proposed accelerator using MobileNetV2 and Transformer for image captioning, and ResNet-18 for image classification. 
We used the same baselines and evaluated them under the same methods and experimental conditions as described in Section~\ref{sec:standalone_analysis}.

\textbf{Comparison of execution time: }We observed that the All-rounder accelerator and baseline accelerators achieve the shortest execution time when the MobileNetV2 and ResNet-18 models are processed in parallel on two 64$\times$128 subarrays. 
Accordingly, we report the execution times based on this structural configuration. 
Our proposed accelerator requires 30.30 ms, while the accelerators based on~\cite{sara} and~\cite{data_flow_mirror} require 33.33 ms and 93.65 ms, respectively.
Due to the low MAC utilization in depthwise convolutions, the baselines require 6.01\% and 60.60\% longer execution times compared to the All-rounder.
The TPU-like SA, constrained by bottlenecks from narrow output buses of the MAC arrays, takes 1.05 sec to complete the same task.

\subsection{Comparison to GPU}\label{sec:gpu_analysis}
\textbf{Baselines and methodology: }Thanks to the widespread use of GPUs for AI applications, we compared the performance and energy efficiency of the training process with a high-end GPU device, i.e., NVIDIA RTX 3090~\cite{nvidia_Rtx}. 
We trained five representative CNN models, which are AlexNet~\cite{alexnet}, VGG16~\cite{vggnet}, ResNet-18~\cite{resnet}, MobileNetV1~\cite{mobilenet_v1}, and DenseNet-121~\cite{densenet}, using a batch size of 128 on the ImageNet dataset~\cite{imagenet}. 
The models were trained using mixed precision on the GPU device as described in~\cite{mixed_training}. 
To measure the power consumption of each AI benchmark, we set up our experimental environment with the automatic mixed precision (AMP) tool from NVIDIA's Apex extension~\cite{nvidia_amp}. 
Additionally, the execution time for a single training iteration was recorded using a built-in Python function.
Given that the GPU device is fabricated using 8nm technology, we scaled the reported values of the All-rounder to align with this technology node for a fair comparison, based on the methodology in~\cite{scaling_factor}.
We compare the performance and energy efficiency of the GPU device with the All-rounder accelerator, which trains the AI benchmarks using bfloat16.

\begin{table}[]
\centering
\caption{Comparisons of performance and energy efficiency between GPU (NVIDIA RTX 3090) device and the All-rounder}
\label{tab:gpu_comp}
\scalebox{0.76}{
\begin{tabular}{|c|c|r|r|r|r|r|}
\hline
\multicolumn{2}{|c|}{\textbf{DNN Benchmark}}                                                                                           & \multicolumn{1}{c|}{\textbf{AlexNet}} & \multicolumn{1}{c|}{\textbf{VGG16}} & \multicolumn{1}{c|}{\textbf{ResNet}} & \multicolumn{1}{c|}{\textbf{MobileNet}} & \multicolumn{1}{c|}{\textbf{DenseNet}} \\ \hline\hline
\multirow{3}{*}{\textbf{\begin{tabular}[c]{@{}c@{}}GPU\\ (\texttt{FP16}\\ +\texttt{FP32})\end{tabular}}}        & \textbf{Runtime {[}ms{]}} & 46.0                                  & 296.4                             & 71.4                                 & 65.9                                    & 214.0                                  \\ \cline{2-7} 
                                                                                              & \textbf{Power {[}W{]}}    & 207.7                                 & 326.7                             & 321.4                                & 322.7                                   & 336.2                                  \\ \cline{2-7} 
                                                                                              & \textbf{GFLOPS/W}         & 41.1                                  & 61.0                              & 36.3                                 & 9.8                                     & 15.5                                   \\ \hline
\multirow{3}{*}{\textbf{\begin{tabular}[c]{@{}c@{}}All-rounder\\ (\texttt{bfloat16})\end{tabular}}}          & \textbf{Runtime {[}ms{]}} & 71.3                                 & 463.7                            & 229.4                                & 256.7                                    & 275.9                                  \\ \cline{2-7} 
                                                                                              & \textbf{Power {[}W{]}}    & 5.3                                  & 5.3                              & 5.3                                 & 5.3                                    & 5.3                                   \\ \cline{2-7} 
                                                                                              & \textbf{GFLOPS/W}         & 2848.3                                 & 6591.5                             & 1880.4                                & 419.0                                   & 2089.4                                  \\ \hline
\end{tabular}}
\end{table}

\textbf{Performance and energy efficiency: }Table~\ref{tab:gpu_comp} presents a summary of the experimental results. 
Since the GPU card is equipped with 20,992 FP16 multipliers and operates at a higher clock frequency (i.e., 1.8 GHz)~\cite{nvidia_Rtx}, the All-rounder accelerator shows 2.30$\times$ lower performance on average compared to the GPU. 
However, due to its significantly lower power consumption, the All-rounder accelerator realizes an improvement in energy efficiency by 81.4$\times$ on average.

\section{Conclusion}
Due to the limited computational capacity of edge and mobile devices, AI applications are offloaded to datacenters. 
ASICs deployed in datacenters receive requests for various applications from users, with each request that may have different requirements in terms of service types (i.e., inference and training), data formats, data dimensions, and multi-tenancy.
To deliver quick responses to users while minimizing hardware costs, these ASICs must efficiently support AI operations under the diverse requirements.
To meet this challenge, All-rounder offers two key advantages: i) flexibility in data format support and ii) high MAC utilization using a morphable MAC array. 
To achieve bit-width flexibility, we propose an area-efficient all-in-one multiplier that can support diverse data formats. 
We also devised a mapping strategy and a flexibly fusible and fissionable MAC array which ensures high resource utilization regardless of operation types, data dimensions, or multi-tenant scenarios.
To validate the effectiveness of our design, we extensively compared not only the multiplier unit itself but also the All-rounder accelerator with various baselines at different operating scenarios.

\section*{Acknowledgment}
This work was partially supported by the Institute of Information \& Communications Technology Planning \& Evaluation (IITP) funded by the Ministry of Science and ICT under Grant RS-2023-00229849; in part by the National Research Foundation of Korea (NRF) funded by the Ministry of Science and ICT under Grant RS-2023-NR077140.
The EDA tool was supported by the IC Design Education Center (IDEC) in South Korea.

\bibliographystyle{plain}
\bibliography{refs}
\vspace{-2mm}
\begin{IEEEbiography}[{\includegraphics[width=1in,height=1.25in,clip,keepaspectratio]{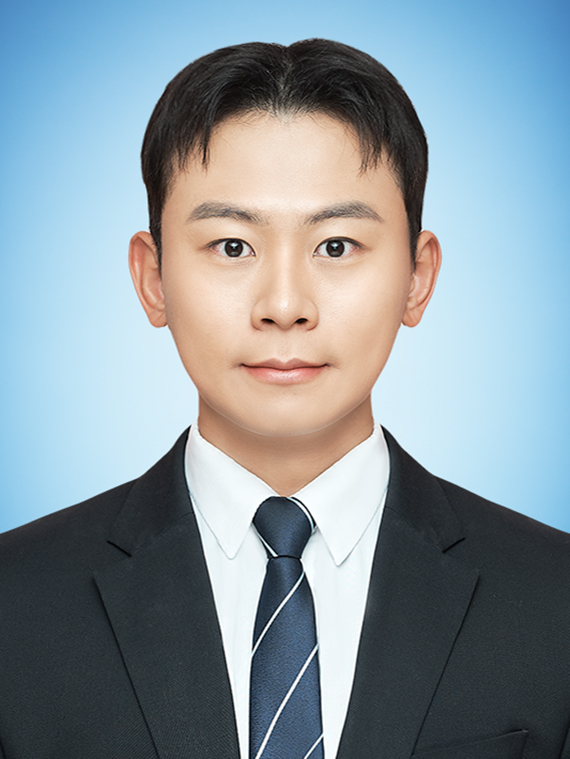}}]{Seock-Hwan Noh}
(S'19) received the B.S. degree in Electrical Engineering from Ulsan National Institute of Science and Technology (UNIST), Ulsan, South Korea, in 2019. 
He is currently pursuing the integrated Master's \& Ph.D degree in Electrical Engineering and Computer Science from Daegu Gyeongbuk Institute of Science and Technology (DGIST), Daegu, South Korea. 
His current research interests include AI training accelerators, circuit design for security and SoC design for autonomous driving cars.
\end{IEEEbiography}\vspace{-2mm}

\begin{IEEEbiography}[{\includegraphics[width=1in,height=1.25in,clip,keepaspectratio]{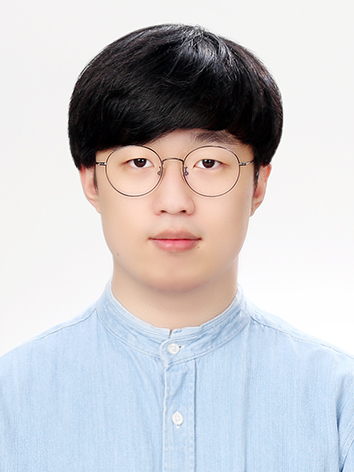}}]{Seungpyo Lee}
(S'21) received the B.S. and M.S. degrees in Electrical Engineering and Computer Science from Daegu Gyeongbuk Institute of Science and Technology (DGIST), Daegu, South Korea, in 2021 and 2024, respectively. 
Currently, he is a data analytics engineer at Fitogether, Seoul, South Korea. 
His research interests include large language models and data quantization. 
\end{IEEEbiography}\vspace{-2mm}

\begin{IEEEbiography}[{\includegraphics[width=1in,height=1.25in,clip,keepaspectratio]{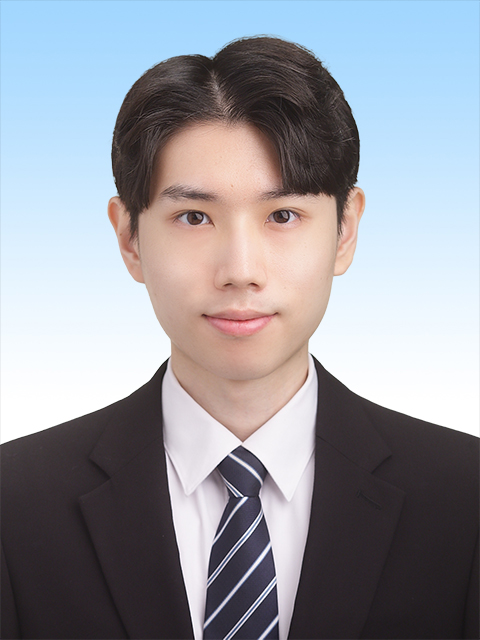}}]{Banseok Shin}
(S'21) received the B.S. degree in Electronic and IT Media Engineering from Seoul National University of Science and Technology, Seoul, South Korea, in 2021 and M.S. degree in Electrical Engineering and Computer Science from Daegu Gyeongbuk Institute of Science and Technology (DGIST), Daegu, South Korea, in 2024.
Currently, he is a circuit design engineer at Samsung Electronics, Suwon, South Korea. 
His research interests include AI training accelerators and circuit design for compute-in-memory.
\end{IEEEbiography}\vspace{-2mm}

\begin{IEEEbiography}[{\includegraphics[width=1in,height=1.25in,clip,keepaspectratio]{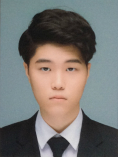}}]{Sehun Park}
(S'18) received the B.S. degree from Daegu Gyeongbuk Institute of Science and Technology (DGIST), Daegu, South Korea, in 2018.
He is currently pursuing the integrated Master's \& Ph.D degree in Electrical Engineering and Computer Science from Daegu Gyeongbuk Institute of Science and Technology (DGIST), Daegu, South Korea.
His current research interests include deep learning training accelerators and 
pruning on neural networks.
\end{IEEEbiography}\vspace{-2mm}

\begin{IEEEbiography}[{\includegraphics[width=1in,height=1.25in,clip,keepaspectratio]{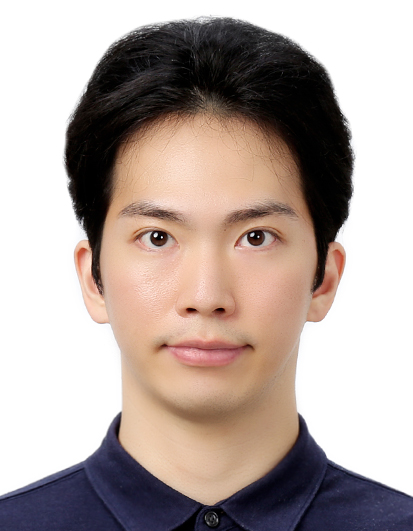}}]{Yongjoo Jang}
(S'19) received the B.S. degree from Daegu Gyeongbuk Institute of Science and Technology (DGIST), Daegu, South Korea, in 2019.
He is currently pursuing the integrated Master's \& Ph.D degree in Electrical Engineering from Korea University, Seoul, South Korea. 
His current research interests include machine learning architecture and AI accelerators.
\end{IEEEbiography}\vspace{-2mm}

\begin{IEEEbiography}[{\includegraphics[width=1in,height=1.25in,clip,keepaspectratio]{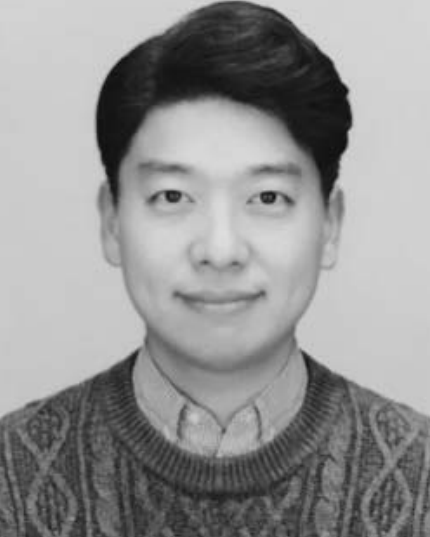}}]{Jaeha~Kung}
(S’13–M’17) received the B.S. degree in Electrical Engineering from Korea University, Seoul, South Korea, in 2010, the M.S. degree in Electrical Engineering from the Korea Advanced Institute of Science and Technology (KAIST), Daejon, South Korea, in 2012, and the Ph.D. degree in Electrical and Computer Engineering from Georgia Institute of Technology, Atlanta, GA, USA, in 2017. 
He is currently an Associate Professor at the School of Electrical Engineering, Korea University, Seoul, South Korea. His current research interests include digital accelerators for emerging applications, efficient AI, system design for artificial intelligence, approximate computing, and low-power VLSI design. He also served as a Technical Program Committee for IEEE/ACM DAC, IEEE/ACM ISLPED, IEEE ISCAS, and IEEE AICAS.

\end{IEEEbiography}\vspace{-2mm}

\end{document}